\documentclass[%
 aip,
 amsmath,amssymb,
 reprint,%
]{revtex4-1}

\usepackage{graphicx}
\usepackage{dcolumn}
\usepackage{bm}
\usepackage[utf8]{inputenc}
\usepackage[T1]{fontenc}
\usepackage{mathptmx}
\usepackage{etoolbox}
\usepackage[version=4]{mhchem}

\makeatletter
\def\@email#1#2{%
 \endgroup
 \patchcmd{\titleblock@produce}
  {\frontmatter@RRAPformat}
  {\frontmatter@RRAPformat{\produce@RRAP{*#1\href{mailto:#2}{#2}}}\frontmatter@RRAPformat}
  {}{}
}%
\makeatother
\begin{document}

\preprint{AIP/123-QED}

\title[A Segmented Heater-Driven, Low-Loss, Reconfigurable Photonic Phase-Change Material-Based Phase Shifter]{A Segmented Heater-Driven, Low-Loss, Reconfigurable Photonic Phase-Change Material-Based Phase Shifter}
\author{Ranjeet Dwivedi}
\email{ranjeetdwivedi2@gmail.com}
\affiliation{ %
INSA Lyon, Ecole Centrale de Lyon, CNRS, Université Claude Bernard Lyon 1, CPE Lyon, INL, UMR5270, 69621 Villeurbanne, France
}%
 
\author{Agraj Yadav}%

\affiliation{%
Univ. Grenoble Alpes, Univ. Savoie Mont Blanc, CNRS, Grenoble INP, CROMA, 38000 Grenoble, France 
}%

\author{Regis Orobtchouk}
\affiliation{ %
INSA Lyon, Ecole Centrale de Lyon, CNRS, Université Claude Bernard Lyon 1, CPE Lyon, INL, UMR5270, 69621 Villeurbanne, France
}%
\author{Benoit Charbonnier}
\affiliation{ %
CEA-LETI, Grenoble, France 
}%
\author{Stephane Malhouitre}
\affiliation{ %
CEA-LETI, Grenoble, France 
}%
\author{Pierre Noé}
\affiliation{ %
CEA-LETI, Grenoble, France 
}%
\author{Fabio Pavanello}%
\email{fabio.pavanello@cnrs.fr}
\affiliation{%
Univ. Grenoble Alpes, Univ. Savoie Mont Blanc, CNRS, Grenoble INP, CROMA, 38000 Grenoble, France 
}%

\date{\today}

\begin{abstract}
Phase-change material (PCM)-based non-volatile multilevel phase shifters are key components in photonic integrated circuits. Electrically, multiple phase levels can be encoded by controlling the heater power and employing different microheater architectures to induce varying degrees of PCM amorphization. However, encoding a large number of levels is not straightforward. In this work, we first investigate a phase shifter structure based on a GeSe PCM integrated on top of a silicon-on-insulator waveguide, employing a simple rectangular-shaped heater under pulse-width modulation (PWM). We numerically demonstrate that multilevel phase shifts can be achieved because of non-uniform heating in the GeSe PCM layer. However, the resulting phase levels for this basic configuration are highly non-linear because of the uniform power dissipation along the light propagation direction characterized by the same cross-section. To overcome this limitation, we designed a novel PCM-based phase shifter with a segmented heater whose width gradually increases along the light propagation direction. This configuration enables the encoding of hundreds of well-spaced phase levels between 0 and $\pi$, facilitated by smoother amorphization arising from the combined effects of non-uniform heating across segments and within each segment, while achieving an insertion loss of only 0.6 dB in the worst case. Furthermore, when evaluating both heater architectures under pulse amplitude modulation (PAM) at a fixed pulse duration, we observe behavior consistent with the trends observed for PWM, confirming the superior performance of the segmented heater design.
\end{abstract}

\maketitle

\section{\label{sec:level1}Introduction}

Reconfigurable photonic integrated circuits (PICs) have emerged as key enablers for a multitude of applications, from sensing and microwave processing to co-packaged optics and computing \cite{Ahmed2025Universal,Bogaerts2020Programmable,Rogers2021Universal3D,Marpaung2019Integrated,Sun2015SingleChip}. One of the central building blocks often used to achieve reconfigurability is the Mach-Zehnder Interferometer (MZI), in which an optical path delay is created between the two arms of the interferometer using a phase shifter, allowing the modification of the proportion of light entering a given port \cite{Bogaerts2020Programmable}. For example, in coherent photonic computing architectures, information is encoded in the amplitude of optical signals, and MZIs meshes that act as complex-valued optical weights are used to perform matrix–vector multiplications (MVMs) \cite{Shen2017Deep}. The performance of these circuits critically depends on the properties of the phase shifters, which can be broadly categorized into volatile and non-volatile states. Phase shifters based on thermo-optic or free-carrier electro-optic effects are volatile in nature and thus require continuous power to maintain their programmed state, leading to high static power consumption. In contrast, non-volatile phase shifters enable energy-efficient reconfiguration by retaining their programmed state without additional power. This characteristic makes non-volatile phase shifters highly attractive for large-scale programmable photonic circuits, including photonic neural networks (PNNs).

Recently, phase-change materials (PCMs) have been explored for non-volatile phase shifting in integrated photonics \cite{Chen2023Nonvolatile,Sun2025Microheater,Rios2022Ultracompact}. Phase shifts in PCMs integrated waveguides can be induced by leveraging the large change in refractive index by switching between crystalline and amorphous states, approximately 2 and 3 orders of magnitude larger than thermo-optical and free-carrier effects, respectively \cite{Brunner2025Roadmap}. This large refractive index modulation significantly reduces the footprint of devices utilizing PCM-based phase shifters compared to those based on electro-optic or thermo-optic effects. However, conventional PCMs such as \ce{Ge2Sb2Te5} (GST) often suffer from high optical losses, which degrade circuit performance  \cite{Albanese2024Optical,Sawant2025HighEndurance}. To address this, alternative low-loss PCMs have been explored, such as \ce{Sb2Se3}, \ce{Sb2S3}, and GeSe, which offer a high refractive index contrast and very low optical absorption \cite{Soref2015ElectroOptical,Delaney2020NewFamily,Teo2022Comparison,Fang2024Arbitrary}. For example, PNNs require the precise implementation of weighted connections; thus, it is essential to realize multilevel phase shifts that can encode a wide range of weight values. Finely-spaced phase levels between the fully crystalline and amorphous PCM states can enable better weight resolution and consequently improve system accuracy. Multilevel phase shifting can be realized in PCM-integrated waveguides by partially amorphizing the PCM through controlled electrical or optical programming \cite{Zhou2022PhaseChangeReview,Xia2024SevenBit,Gong2024SixBit,Dwivedi2025UltraLowLoss}. By modulating the degree of amorphization, intermediate refractive index states can be achieved, resulting in a continuous range of phase shifts. In electrical programming, this control is implemented by varying the energy of the programming pulses \cite{Chen2023Nonvolatile,Sun2025Microheater,Rios2022Ultracompact,Adya2025Interleaved,Wei2023ElectricallyProgrammable,Meng2023ElectricalPRAM}. However, achieving uniform and well-spaced phase levels remains challenging due to the complex and often non-uniform dynamics of the amorphization process \cite{Sun2025Microheater,Brunner2025Roadmap}. Therefore, considerable effort has recently been devoted to optimizing the switching process by adjusting the heating geometry and pulse energy \cite{Chen2023Nonvolatile,Sun2025Microheater,Rios2022Ultracompact,Brunner2025Roadmap,Adya2025Interleaved}.

In this work, we investigate a novel GeSe-based low-loss PCM-integrated phase shifter capable of achieving multilevel phase control. Electrical programming is examined through two approaches: (1) varying the pulse duration at fixed power and (2) varying the power at a fixed pulse duration. Two heater configurations are analyzed: a conventional constant-width heater and a segmented heater with a width that gradually increases along the light propagation direction. Numerical simulations reveal that while the simple constant-width heater under pulse-width modulation (PWM) can encode multiple phase levels, the spacing between these levels is highly irregular because of non-uniform local heating within the GeSe layer and the uniform heating along the light propagation direction. In contrast, the segmented heater configuration enables the encoding of hundreds of well-spaced phase levels, resulting from smoother amorphization of the combined GeSe segments while achieving low optical insertion losses, equal to only 0.6 dB in the worst case. Although individual segments exhibit irregular amorphization, the variation in heating across segments ensures different degrees of amorphization for the same programming conditions, leading to an overall smooth phase change response. These results demonstrate that the segmented heater design is highly suitable for integration into energy-efficient coherent PNNs and, more broadly, for the next generation of programmable photonic systems \cite{Pavanello2023Neuropuls}. 

\section{Device structure}
\begin{figure}[h!]
    \centering
    \includegraphics[width=\linewidth, trim=0cm 2cm 1cm 0cm,clip]{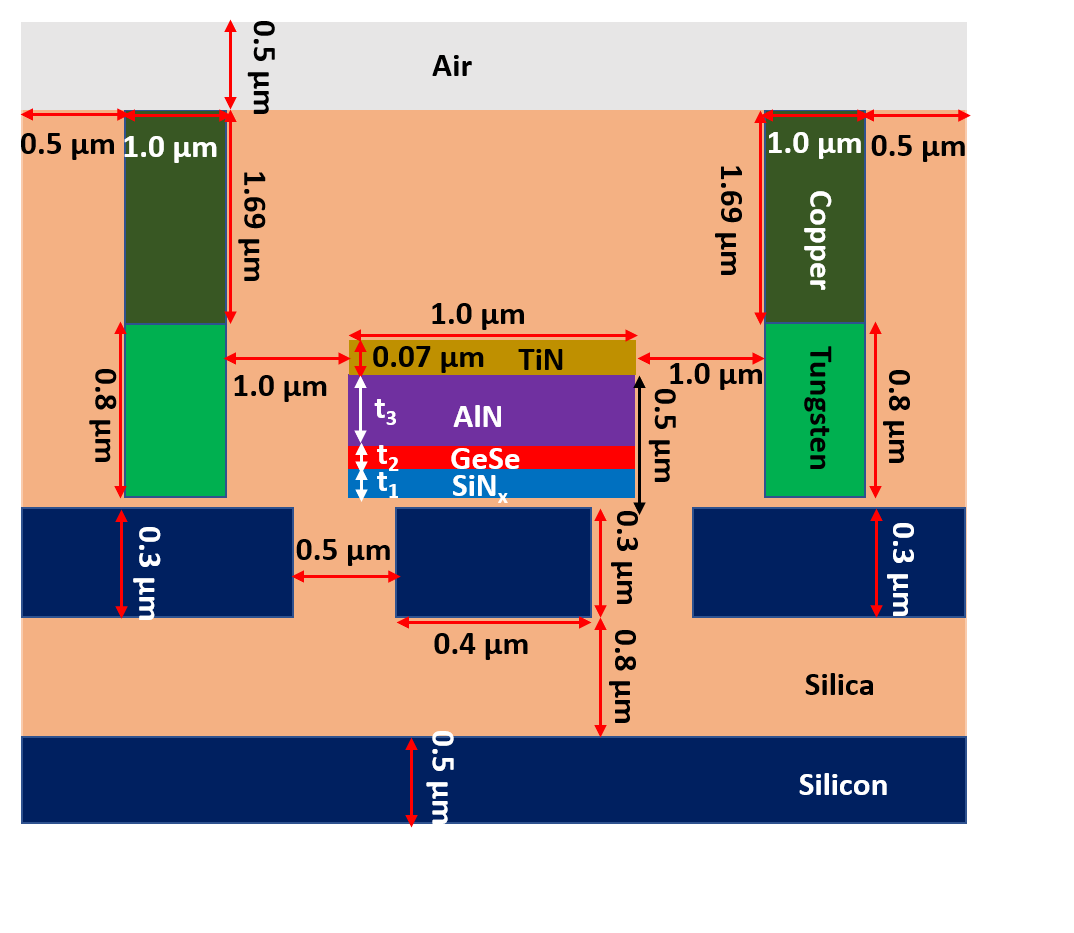} 
    \caption{Schematic of the cross-section of the considered phase shifter structure for the thermal calculations.}
    \label{fig:Fig1}
\end{figure}

The cross-sectional schematic of the phase shifter structure considered is shown in Fig.\ref{fig:Fig1}. The central silicon region consists of the waveguide core on top of which a SiN$_x$ layer is placed, followed by the GeSe phase change material (PCM) layer and an AlN layer to protect the GeSe film from detrimental oxidation and improve heating efficiency. A thin SiO$_2$ gap of 10 nm below the  SiN$_x$ layer and the Si core is present to simulate a chemical-mechanical polishing step. A TiN heater layer is placed 500 nm from the top of the Si core to ensure efficient heating while minimizing the effect on the optical propagation characteristics. The thicknesses of the buried oxide (BOX) layer and the Si substrate are considered as 800 nm and 500 nm, respectively. The silicon slabs on either side of the core are placed at a distance of 500 nm to quickly dissipate the heat to achieve rapid cooling of the GeSe without affecting the optical characteristics. The tungsten and copper vias are placed at a distance of 1 µm from the heater edge in the trench region to further enhance the cooling rate of GeSe. The thickness of the SiO$_2$ cladding is 2.5µm, and a 0.5-µm-thick air region is considered the top material. The thickness and width of the silicon core are taken as 0.3 and 0.4 µm, respectively. The GeSe, SiN$_x$,  AlN, and TiN layers each have a width of 1.0 µm. The thicknesses of the  SiN$_x$, GeSe and AlN layers are denoted as $t_1$, $t_2$ and $t_3$, respectively, while the thickness of TiN is fixed at 70 nm. The thicknesses of the tungsten and copper vias are taken as 0.8 and 1.69 µm, respectively. The widths of the SiO$_2$ buffer in the leftmost and rightmost regions are set at 0.5 µm.

\section{Thermal Simulations and Phase Shift Modelling}
The temperature distribution within the GeSe layer was simulated using Lumerical HEAT\cite{AnsysHeatSolver}. The mesh edge lengths in the simulation were set to a minimum of 1 nm and a maximum of 100 nm, with a triangle quality of 30. The simulation time step ranged from 1 to 10 ns. The initial temperature was set at an ambient value of 20°C. All boundaries, except the upper surface, were kept at ambient temperature. Convection boundary conditions were applied at the glass–air and copper–air interfaces, with a convection coefficient of 5 W/m²K \cite{awbi1998calculation}. In the simulation, the thickness of the SiN$_x$ layer ($t_1$) was set to 20 nm and the GeSe thickness ($t_2$) was chosen to be 35 nm. The thickness of the AlN layer ($t_3$) was set to 435 nm, filling the entire gap between the GeSe layer and the heater layer, enhancing the heating efficiency and cooling rate due to its higher thermal conductivity compared to SiO$_2$. For the proposed configurations, a phase shift $\pi$ is obtained for a phase shifter length of 85.5 µm. The heating was induced by a low-stress TiN heater with a power density of 7.5 mW/µm and a pulse duration of 2 µs. The resistivity of low-stress TiN is assumed to remain constant throughout the temperature range \cite{Creemer2008Microhotplates}. This assumption can be relaxed if the experimental data show a temperature-dependent resistivity, which typically depends on the TiN film quality. In that case, the effective power dissipation can be kept consistent by adjusting the pulse duration or amplitude accordingly. A rise and fall time of 10 ns is considered for the heat pulse. The thermal properties of all materials used in the calculations are summarized in Table \ref{tab:table1}.
\begin{table}[ht]
    \centering
    \caption{Thermal properties of different materials used in the calculations}
	\begin{ruledtabular}
    \begin{tabular}{cccc}  
        Material &Thermal conductivity & Heat capacity & Density \\
        &(W/mK) & (J/kgK) & (kg/m$^3$) \\
        \hline
        TiN\cite{Dwivedi2025UltraLowLoss}   & 29    & 388      & 5210   \\
        AlN   & 90\cite{vaziri2025aln}    & 740\cite{BachrcSi3N4AlNMaterials}      & 3160\cite{drusedau2003properties}   \\
        SiN$_x$\cite{mastrangelo1990thermophysical} & 3   & 700      & 3000   \\
        GeSe\cite{Dwivedi2025UltraLowLoss}  & 1.5   & 329      & 5520   \\
        Si\cite{Dwivedi2025UltraLowLoss}    & 148   & 711      & 2330   \\
        SiO$_2$\cite{Dwivedi2025UltraLowLoss} & 1.38 & 709     & 2203   \\
        Cu \cite{AnsysMaterialDatabase}   & 397   & 385      & 8933   \\
        W\cite{PlanseeTungsten}     & 164   & 130      & 19250  \\
        Air\cite{AnsysMaterialDatabase}   & 0.0263 & 1006.43 & 1.17659 \\
    \end{tabular}
	\end{ruledtabular}
    \label{tab:table1}
\end{table}

\begin{figure}[h!]
    \centering
    \includegraphics[width=\linewidth, trim=0cm 0cm 0cm 0cm,clip]{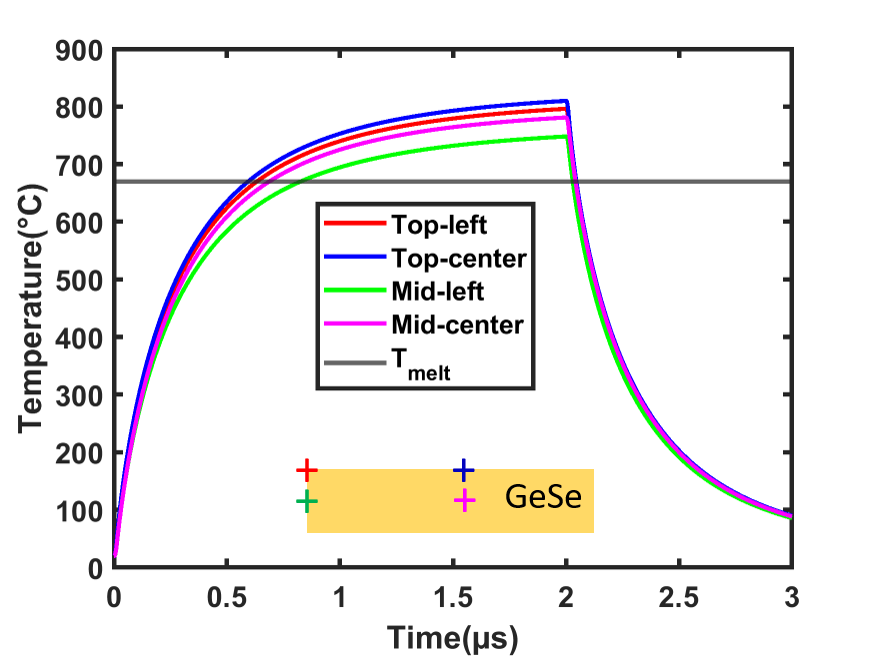} 
    \caption{Temporal variation of temperature at four specific locations within the GeSe. }
    \label{fig:Fig2}
\end{figure}

Figure \ref{fig:Fig2} shows the temporal evolution of the temperature at four specific locations within the GeSe layer, denoted in the inset. The results indicate that different regions of GeSe reach the melting temperature at different times. For instance, the melting threshold is crossed at 594 ns for the top-center point, 630 ns for the top-left, 687 ns for the mid-center, and 824 ns for the mid-left location. This non-uniform heating implies that the phase transition in GeSe does not occur simultaneously across the entire layer; instead, different regions undergo phase change sequentially. Consequently, varying the duration of the heat pulse allows for control over the degree of amorphization. This enables encoding multiple phase-shift levels by simply adjusting the pulse duration at a fixed power, i.e., using pulse width modulation (PWM), which is a simpler approach from a system-driving point of view than pulse amplitude modulation (PAM). This is especially true because of the large driving powers needed for electrical heaters.

\begin{figure}[h!]
    \centering
    \includegraphics[width=\linewidth, trim=0.8cm 0cm 1.1cm 0.4cm,clip]{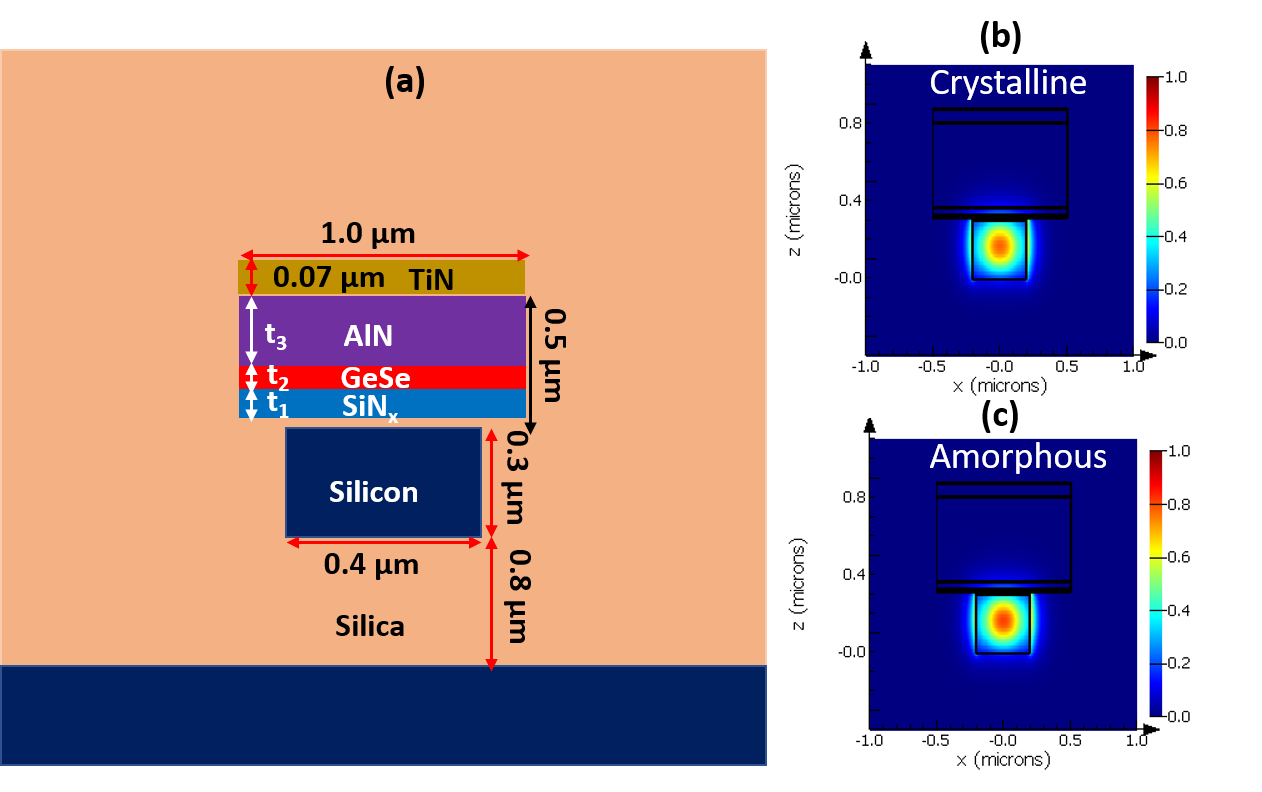} 
    \caption{(a) Schematic of the simulation domain for the calculation of optical propagation characteristics; spatial distribution of electric field intensity in (b) crystalline and (c) amorphous states of GeSe.}
    \label{fig:Fig3}
\end{figure}

Subsequently, we calculated the temperature distribution across a mesh grid within the GeSe layer for various pulse durations, with a time step of 1 ns. The resulting temperature profiles from the thermal simulations were imported into Lumerical FDE to perform optical mode calculations. Our thermal design ensures that the cooling rate from the melting temperature to the glass transition temperature is greater than 1K/ns at every point in the mesh grid where the temperature exceeds the melting point; therefore, a local amorphization is assumed at such points in the mesh grid \cite{Wright2011DesignRewritable,Wuttig2017PhaseChangeReview}.  The simulation domain for mode analysis, illustrated in Fig. \ref{fig:Fig3} (a), is 2-µm-wide and 4-µm-thick. The horizontal mesh size was set to 10 nm, while the vertical mesh size was non-uniform: 2 nm in the SiO$_2$ gap and SiN$_x$, 1 nm in GeSe, 5 nm in AlN and TiN, and 10 nm in the remaining regions. The simulations were carried out at a wavelength of 1.55 µm. The refractive indices of Si, SiO$_2$, and TiN were obtained from Palik's data \cite{Palik1998Handbook}. The refractive index of SiN$_x$ is taken as 2.0 \cite{Luke2015Broadband}, while the values for crystalline and amorphous GeSe are $3.258987+0.00104i$ and 3.085431 \cite{Albanese2024Optical}, respectively. The refractive index of AlN is taken as $2.018+0.00008i$ \cite{beliaev2021thickness}. The spatial distribution of the normalized electric field intensity of the fundamental TE mode in the crystalline and amorphous states of GeSe is shown in Figs. \ref{fig:Fig3}(b) and \ref{fig:Fig3}(c), respectively.

\begin{figure}[h!]
    \centering
    \includegraphics[width=\linewidth, trim=0.8cm 0cm 1.3cm 0.4cm,clip]{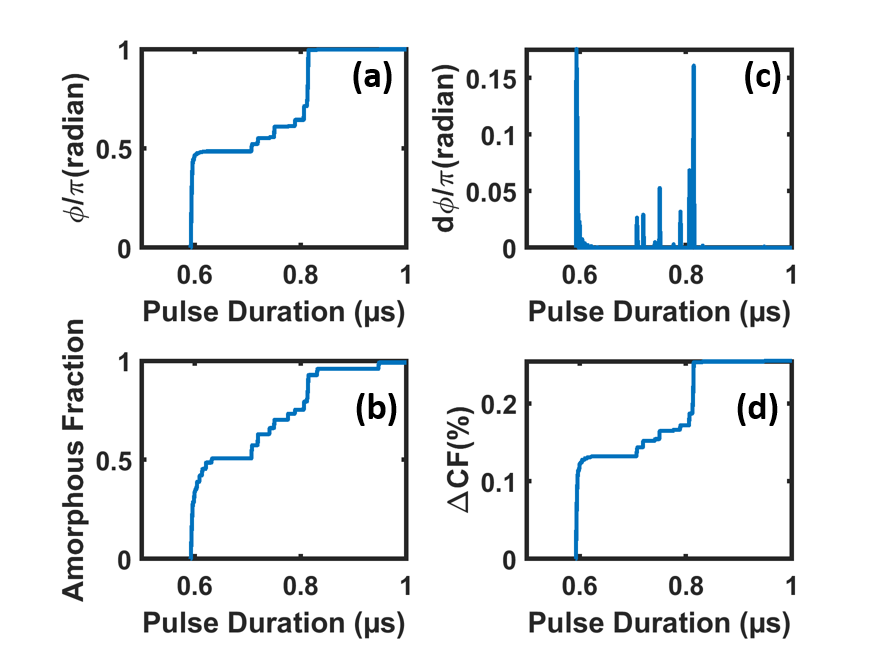} 
    \caption{Variation of (a) phase shift, (b) amorphous fraction, (c) difference between adjacent phase shift levels (d$\phi$), and (d) the change in confinement factor in GeSe ($\Delta$CF), with the pulse duration for a heater with constant-width along the light propagation direction.}
    \label{fig:Fig4}
\end{figure}

The phase shift ($\phi$) is determined from the change in the effective index of the fundamental TE mode as the amorphization fraction varies. Figure \ref{fig:Fig4}(a) shows the variation of the phase shift as a function of the pulse width. The amorphization begins at approximately 593 ns, and as the pulse duration increases, the amorphous fraction of GeSe, calculated by averaging the permittivity distribution across the mesh grid, also increases, as illustrated in Fig. \ref{fig:Fig4}(b). For a pulse duration of 832 ns, a phase shift of $\pi$ is achieved. However, the encoded phase shift levels are highly non-uniform. This is evident from Fig. \ref{fig:Fig4}(c), which presents the difference between adjacent phase shift levels ($d\phi$) for a 1 ns variation in pulse width. At 594 ns, $d\phi$ goes up to 0.175$\pi$ for just a 1 ns increment, indicating abrupt changes in the phase shift. This behavior arises from non-uniform, localized amorphization within the GeSe layer and uniform power dissipation across the light propagation direction (no change in the cross-section along this direction). Moreover, Fig. \ref{fig:Fig4}(d) shows that the variation in phase shift closely follows the change in the confinement factor ($\Delta$CF) of the GeSe layer relative to its fully crystalline state. Such irregular phase encoding as a function of pulse width can severely degrade the performance of PNNs. Therefore, phase shifters based on this design are unsuitable for finely-tuned reconfigurability, e.g., in high-accuracy PNNs when operated under PWM.

\begin{figure}[h!]
    \centering
    \includegraphics[width=\linewidth, trim=1.1cm 0cm 1.3cm 0.4cm,clip]{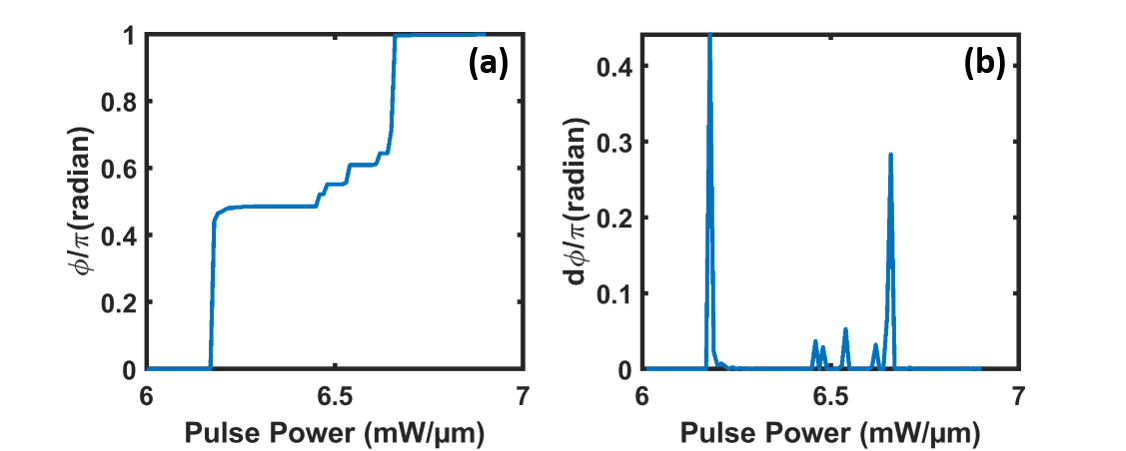} 
    \caption{Variation of (a) phase shift, and (b) difference between adjacent phase shift levels (d$\phi$), with the pulse power}
    \label{fig:Fig5}
\end{figure}

Next, we evaluated the phase shift under PAM driving while keeping the pulse duration fixed at 2 µs. The power was swept from the onset of amorphization to the point where the GeSe layer was fully amorphized, ranging from 6.0 mW/µm to 6.9 mW/µm in increments of 0.01 mW/µm. The resulting encoded phase shift levels, along with the differences between adjacent levels, are presented in Figs. \ref{fig:Fig5}(a) and \ref{fig:Fig5}(b). The results reveal that the phase shift again exhibits a highly non-uniform variation, which is attributed to the irregular amorphization behavior of the GeSe as the heater power increases. The maximum value of $d\phi$ in this case is increased to 0.44$\pi$. Consequently, PAM driving also proves unsuitable for encoding well-spaced phase shift levels. This result is not surprising, as PAM and PWM drivings are correlated by the overall pulse energy provided to the system for pulse durations much shorter than the thermal time constant. 

\section{Segmented Heater Design and Phase Shift Modelling}
To achieve well-spaced phase shift levels, we introduced a segmented heater design with increasing width along the light propagation direction. Increasing the width of the heater reduces its electrical resistance, thus decreasing power dissipation and lowering the local temperature in the GeSe layer. As a result, GeSe segments beneath heaters of different widths experience varying degrees of amorphization. This segmentation approach enables a smoother overall amorphization profile, especially beneficial when considering the non-uniform mode field profile, as the combined response of the segments is more gradual compared to that of a phase shifter with a uniform heater width. 
\begin{figure}[h!]
    \centering
    \includegraphics[width=\linewidth, trim=0cm 3.0cm 0cm 2.5cm,clip]{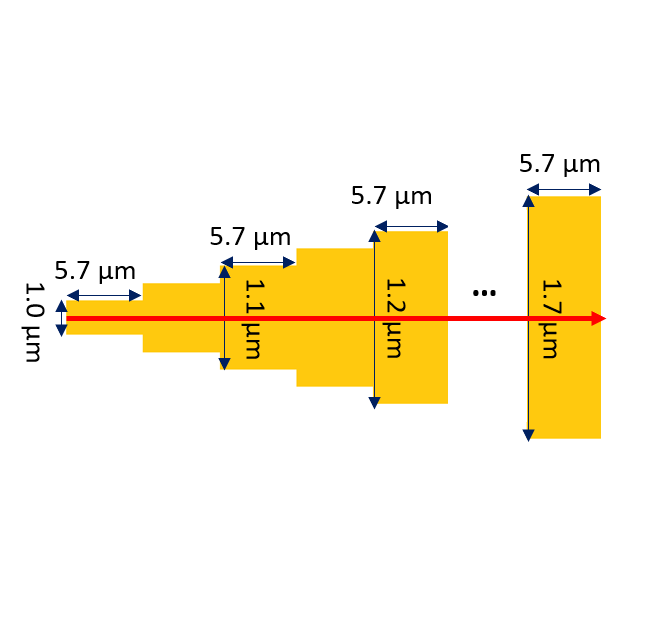} 
    \caption{Schematic of the segmented heater with increasing width along the light propagation direction}
    \label{fig:Fig6}
\end{figure}
\begin{figure*}[t]
    \centering
    \includegraphics[width=\linewidth, trim=0.8cm 0cm 1.3cm 0cm,clip]{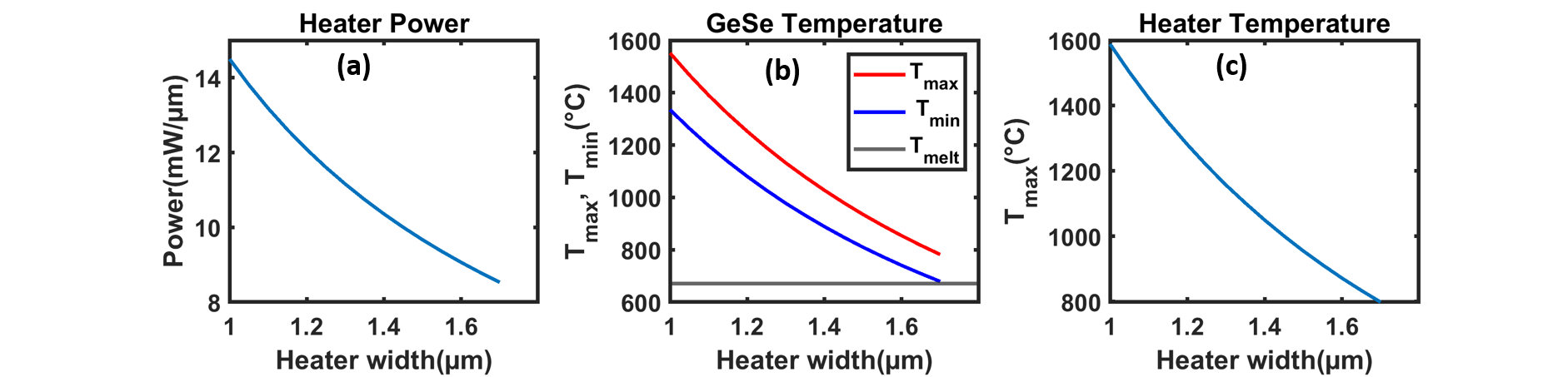} 
    \caption{Variation of (a) heater power, (b) maximum and minimum temperatures across GeSe ($T_{max}$ and $T_{min}$), and (c) maximum temperatures across TiN, with the heater width.}
    \label{fig:Fig7}
\end{figure*}
\begin{figure*}[t]
    \centering
    \includegraphics[width=\linewidth, trim=0.8cm 0cm 1.3cm 0cm,clip]{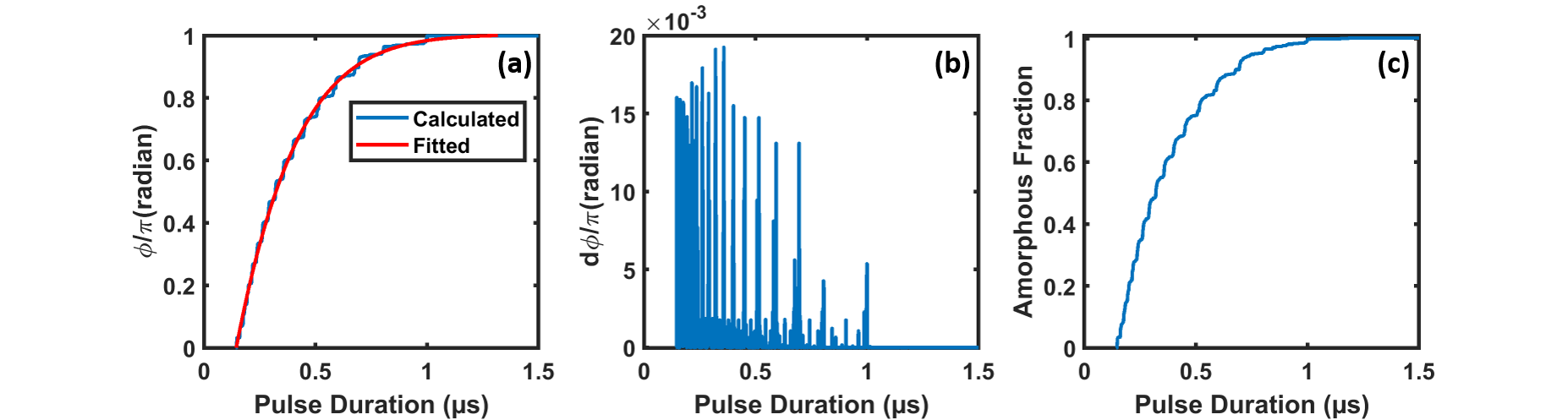} 
    \caption{Variation of (a) phase shift ($\phi$), (b) difference between adjacent phase shift levels (d$\phi$), and (c) amorphous fraction with the pulse duration for the combined GeSe segments.}
    \label{fig:Fig8}
\end{figure*}

\begin{table*}[ht]
    \centering
    \caption{A comparison of encoded phase-shift levels for constant and variable-width architectures}
	\begin{ruledtabular}
    \begin{tabular}{ccccc}
        
        Pulse width & No. of levels & No. of levels & $d\phi$ constant-width & $d\phi$ variable-width \\
        resolution (ns) &constant-width &variable-width & &
        \\
        \hline
        1 & 19 & 110 & $0.004\pi$ to $0.174\pi$ & $0.004\pi$ to $0.019\pi$ \\
        2 & 17 & 81  & $0.004\pi$ to $0.222\pi$ & $0.004\pi$ to $0.037\pi$ \\
        5 & 12 & 61  & $0.005\pi$ to $0.314\pi$ & $0.004\pi$ to $0.056\pi$ \\
        
    \end{tabular}
	\end{ruledtabular}
    \label{tab:table2}
\end{table*}

For a fixed increase in the heater width, as the number of heater segments increases, the difference in the power dissipations between the first and last segments increases, which enhances the heater temperature difference. We selected the number of segments such that the GeSe layer beneath the last heater segments can be fully amorphized while the temperature in the first heater segment is below the SiO$_2$ melting temperature, to avoid the melting of SiO$_2$ near the TiN heater.  For our calculations, we considered 15 heater segments with widths ($w$) increasing from 1.0 µm to 1.7 µm in increments of 0.05 µm, as shown in Fig. \ref{fig:Fig6}. The temperature distribution across the GeSe segments is then calculated.

Figure \ref{fig:Fig7}(a) presents the variation in the heater power as a function of the heater width. The highest power dissipation, 14.5 mW/µm, occurs for $w$ = 1.0 µm, while the lowest, 8.53 mW/µm, corresponds to $w$ = 1.7 µm. Variations in maximum and minimum GeSe temperatures in different segments are shown in Fig. \ref{fig:Fig7}(b). For widths up to 1.7 µm, the minimum temperature remains above the melting threshold, indicating complete amorphization of all these segments. Furthermore, Fig. \ref{fig:Fig7}(c) shows that the maximum heater temperature (1588 °C) remains below the melting point of SiO$_2$ (1713 °C), confirming that there is no melting of nearby SiO$_2$. 

Next, we evaluated the phase shift for the segmented heater by varying the heat pulse duration (PWM driving), similarly to the approach used for the constant-width heater. Each segment has a length of 5.7 µm, giving a total phase shifter length of 85.5 µm. The variation of the overall phase shift with pulse duration is shown in Fig. \ref{fig:Fig8}(a). Unlike the constant-width configuration, the segmented design exhibits a smooth phase shift response, as indicated by the small differences between adjacent phase shift values $d\phi$ in Fig. \ref{fig:Fig8}(b), despite the individual segments showing highly non-uniform behavior. The maximum $d\phi$ is approximately 0.019$\pi$ at $\Delta$t = 358 ns, while for most pulse durations it remains very small. This smooth behavior is attributed to the gradual and averaged amorphization across multiple segments, as illustrated in Fig. \ref{fig:Fig8}(c), in contrast to the abrupt amorphization observed with a single heater segment. With this configuration, a phase shift ranging from 0 to $\pi$ is achieved for pulse durations between 145 and 1250 ns. Although the spacing between phase shift levels is not perfectly uniform, it is sufficiently consistent to map a large number of phase shift levels between 0 and $\pi$ to the corresponding pulse widths. 

Table \ref{tab:table2} shows a comparison of the number of encoded phase-shift levels for different pulse width resolutions. In each case, the spacing between adjacent phase shifts is considered to be not less than 0.04$\pi$. As can be seen, the number of encoded levels in the case of variable heater width is 110, 81, 61, compared to 19, 17, 12 for constant width at 1 ns, 2 ns, and 5 ns resolutions in pulse width. Moreover, the range of $d\phi$ is quite small in the former case. Therefore, the segmented heater with variable width along the light propagation direction provides a significantly smoother phase-shift response under PWM driving than the constant-width heater, with the possibility of encoding hundreds of well-spaced phase-shift levels.

The insertion loss is found to vary from 0.6 dB to 0.38 dB for the phase shift varying from 0 to $\pi$. The phase shift variation can be fitted with an error function, which is displayed in Fig. \ref{fig:Fig8}(a), showing a good fit with an R-square value of 0.9987. The fitted function is $y=a\operatorname{erf}(b(x-c))+d$, where $y$ is the phase shift, $x$ is the pulse duration in microseconds, $a$=22.2106, $b$=1.0968, $c$=-1.1464 and $d$=-21.2076.

\begin{figure}[h!]
    \centering
    \includegraphics[width=\linewidth, trim=1.2cm 0cm 1.6cm 0.4cm,clip]{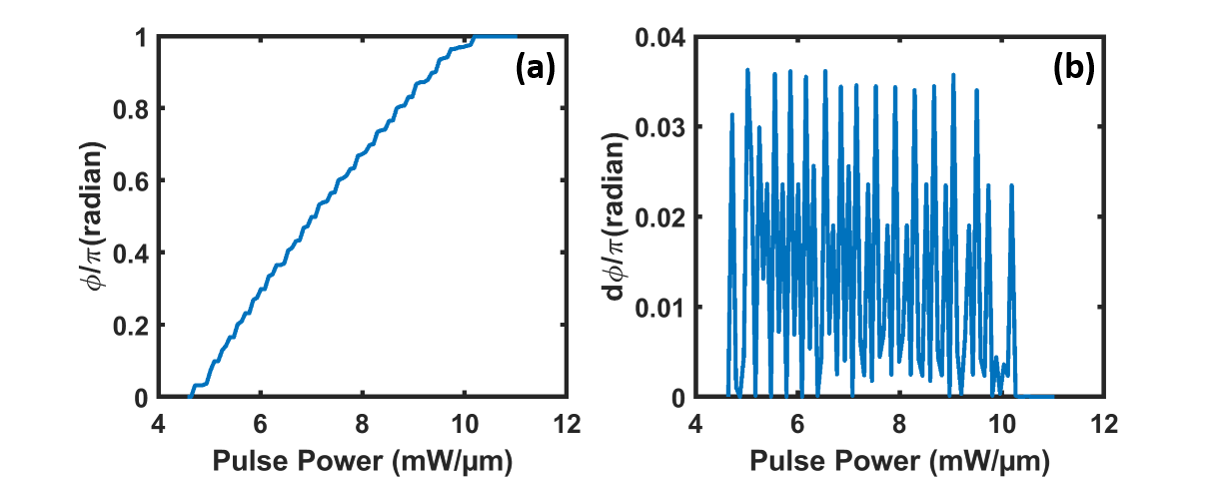} 
    \caption{Variation of (a) phase shift, and (b) difference between adjacent phase shift levels (d$\phi$), with the pulse power for the combined GeSe segments.}
    \label{fig:Fig9}
\end{figure}

We also calculated the phase shift for the variable-width heater configuration under PAM driving, with a fixed pulse duration of 2 µs. The power dissipation in the first heater segment was swept from 6.0 mW/µm, corresponding to the onset of amorphization in the first GeSe segment, to 14.5 mW/µm, corresponding to complete amorphization of the final GeSe segment, in intervals of 0.1 mW/µm. The average power dissipation across all segments therefore varies from 4.56 mW/µm to 11.03 mW/µm.  As expected, the phase shift exhibits similar behavior to that under PWM driving, as indicated by a smoother phase shift variation, illustrated in Figs. \ref{fig:Fig9}(a) and \ref{fig:Fig9}(b). These results confirm that the segmented heater design provides significantly improved uniformity in phase shift variation under both PWM and PAM driving. This improved performance makes the segmented design more suitable for integration into Mach-Zehnder interferometers to enable precise reconfigurability of phase shifters (limited by the bit resolution of digital-to-analog converters (DAC)) and related PICs, e.g., for high-accuracy PNNs\cite{Hejda2024Neuromorphic}. 

Our design is consistent with typical foundry design rules. Consequently, the heater width is set to a minimum of 1 µm and increased in steps of 50 nm. The average power dissipation across all segments varies between 0.39 W and 0.94 W, corresponding to different phase shift levels between the fully crystalline and fully amorphous states of GeSe. Although narrower heaters could further reduce total power consumption, our design is particularly suited for applications requiring a low rate of reconfigurability, where weight programming dominates the overall energy cost. Moreover, the proposed segmented heater design can be extended to other waveguide platforms and phase-change materials, whilst potentially reducing the required power consumption by modifying the currently used foundry design rules.

\section{Conclusions}
We investigated a GeSe-integrated silicon waveguide phase shifter incorporating a TiN heater. Our analysis revealed that non-uniform local heating leads to spatially varying amorphization in GeSe, causing different regions to switch at different pulse durations. As a result, multiple phase-shift levels can be encoded by varying the pulse duration at constant power (PWM driving). However, these levels are highly irregular because of abrupt and uneven amorphization. To address this, we proposed a segmented heater design with gradually increasing width along the light propagation direction. This architecture enabled a smoother overall amorphization across GeSe segments, producing a more continuous phase shift response. With this approach, hundreds of well-spaced phase-shift levels between 0 and $\pi$ can be mapped to different pulse durations. Under PAM driving, both heater designs exhibited similar trends as under PWM driving. This study demonstrates that segmented heaters with variable width provide superior performance compared to uniform heaters, enabling energy-efficient, multilevel, non-volatile phase shifters for next-generation programmable photonic systems, particularly photonic neural networks. 

\section*{Data Availability Statement}
Data supporting the findings of this study are available
from the corresponding author on a reasonable request.
\begin{acknowledgments}
This work received funding from the Horizon Europe research and innovation program of the European Union under Grant Agreement No. 101070238. However, views and opinions expressed are those of the author(s) only and do not necessarily reflect those of the European Union. Neither the European Union nor the granting authority can be held responsible for them.
\end{acknowledgments}
\section*{AUTHOR DECLARATIONS}
\subsection*{Conflict of Interest}
The authors have no conflicts of interest to disclose.
\section*{References}
\bibliography{Ref}
\end{document}